\documentclass[adp,fleqn]{w-art}
\usepackage{times}
\usepackage[]{graphicx}
\usepackage{amsmath,amssymb}
\chardef\bslash=`\\ % p. 424, TeXbook

\begin{document}
%%
%%  Most of the following commands will be completed by the publisher.
%%
%%  The copyrightyear is defined in the .clo file as the first argument
%%  of the copyrightinfo command. If the copyrightyear differs from that
%%  value it might be adjusted by the following definition:
%%
%% \renewcommand{\copyrightyear}{2002}% uncomment to change the copyrightyear.
%%
\DOIsuffix{theDOIsuffix}
%%
%% issueinfo for header and copyright line
%%
\Volume{12}
\Issue{1}
\Copyrightissue{01}
\Month{01}
\Year{2003}
%%
%%  First and last pagenumber of the article. If the option
%%  'autolastpage' is set (default) the second argument may be left empty.
\pagespan{1}{}
%%
%%    Dates will be filled in by the publisher. The 'reviseddate' and
%%    'dateposted' (Published online) entry may be left empty.
%\Receiveddate{15 November 1900}
%%
%% \Reviseddate{30 November 1900}
%%
%\Accepteddate{2 December 1900}
%% \Dateposted{3 December 1900}
%%
%%

%%%%%%%%%%%%%%% start working

\title{Nonlinear Transport through Quantum Dots Studied by the Time-Dependent DMRG}

%%   Information for the first author.
\author[S.\ Kirino]{Shunsuke Kirino\footnote{Corresponding
     author \quad E-mail: {\sf kirino@issp.u-tokyo.ac.jp}, Phone: +81\,471\,363\,272}}
\address{The Institute for Solid State Physics, The University of Tokyo,
     Kashiwanoha 5-1-5, Kashiwa, Chiba 277-8581 Japan}
%%
%%    Information for the second author
\author[K. Ueda]{Kazuo Ueda}
\begin{abstract}
 Recent developments on studies of transport through quantum dots
 obtained by applying the time-dependent density matrix renormalization
 group method are summarized.
 Some new aspects of Kondo physics which appear in nonequilibrium steady
 states are discussed both for the single dot case and for the serially
 coupled double-quantum-dot case.
\end{abstract}
\maketitle

%% If there is not enough space inside the running head
%% for all authors including the title you may provide
%% the leftmark in one of the following three forms:

%% \renewcommand{\leftmark}
%% {F.\ Author: Short Title}

%% \renewcommand{\leftmark}
%% {F.\ Author and S.\ Author: Short Title}

%% \renewcommand{\leftmark}
%% {F.\ Author et al.: Short Title}

%% \tableofcontents  % Produces the table of contents.
\graphicspath{{figures/}}

\section{Introduction}
\label{sect1}
Advancements in microfabrication technology provide us
opportunities to investigate transport properties through quantum dots.
Stimulated by possible application to quantum computation, number of
interesting systems are now being produced by arranging multiple
quantum dots in various geometries.

In the simplest case, namely a single quantum dot system, it was
predicted \cite{linear_1} and experimentally observed \cite{exp_1} that the linear
conductance is enhanced at low temperatures up to the unitary limit by
the Kondo effect when odd number of electrons are accommodated in the
dot.

To measure a current through quantum dots a finite bias voltage is
applied between source and drain electrodes.
Therefore, in principle, we always confront the problem of nonlinear
transport.
Away from the linear response regime, the differential conductance for
single dot system is quickly suppressed reflecting the sharp Kondo
resonance peak at the Fermi energy.
The nonlinear transport through quantum dots defines a prototypical
problem for electron correlation under a nonequilibrium condition.
Various theoretical approaches have been applied to this problem but our
understanding of the problem is still far from complete.
Therefore it is desirable to obtain unbiased results by using some
computational approaches.

Recently there has been progress in computational physics for strongly
correlated electron systems from two complementary approaches.
One is from the infinite dimension initiated by Metzner and
Vollhardt \cite{MetznerVollhardt},
which was later developed into the dynamical mean field theory (DMFT) \cite{DMFT_1}.
It is interesting to note that the Kondo effect is relevant to the DMFT
since in the infinite dimension the self-energy becomes local, which
is determined by similar physics as the Kondo effect.
The other is the density matrix renormalization group method developed
by White \cite{DMRG}, which is an approach from one dimension.
Among many numerical techniques which have been extended to nonequilibrium
situations, the time-dependent density matrix renormalization group
method (TdDMRG) \cite{tdDMRG_1} has proven to be a powerful tool to
tackle the nonlinear transport in strongly correlated electron systems,
such as the single quantum dot system
\cite{Al-Hassanieh,KirinoQD,KirinoQD2,Heidrich-Meisner} and the
interacting resonant level model \cite{Schmitteckert1,Schmitteckert2}.
In this paper we discuss our recent activities to understand the
nonequilibrium transport through quantum dots by using the TdDMRG.
We will briefly summarize our results concerning the single quantum dot
\cite{KirinoQD,KirinoQD2} and then show recent results on the serial
double-dot system.

\section{Single Quantum Dot System}
We consider a system consisting of a quantum dot and two leads.
We focus on the situation where the level spacing in the quantum dot is
large compared to the other energy scales in the system, so that only
one level in the dot is active and the others are effectively frozen.
Concerning the leads we assume that single channel is relevant to
transport, which can be described by the 1-D tight-binding model.
The Hamiltonian for the system is expressed as follows:
\begin{align}
\label{qdot_Hamiltonian}
 H(\tau)=
&-t \sum_{i <-1} \sum_{\sigma} ( c_{i \sigma}^{\dagger} c_{i+1 \sigma} + h.c.)
 -t \sum_{i > 0} \sum_{\sigma} ( c_{i \sigma}^{\dagger} c_{i+1 \sigma} + h.c.) \notag \\
&-t' \sum_{\sigma} \left[
   (c_{-1 \sigma}^{\dagger} c_{0 \sigma} + h.c.)
 + (c_{ 0 \sigma}^{\dagger} c_{1 \sigma} + h.c.) \right]  \notag \\
&-\frac{U}{2} \sum_{\sigma} c_{0 \sigma}^{\dagger} c_{0 \sigma}
 + U c_{0 \uparrow}^{\dagger} c_{0 \downarrow}^{\dagger} c_{0 \downarrow} c_{0 \uparrow}
 +\frac{eV}{2} \theta(\tau) \left(N_L - N_R\right),
\end{align}
where $\tau$ is the time variable, the quantum dot is located at the
$0$th site and $c_{i\sigma}^{\dagger}$ creates an electron with spin
$\sigma$ at $i$th site.
$t, t',U$ and $V$ are the hopping amplitude in leads, the one between
the dot and the neighboring sites, the Coulomb repulsion energy and the
voltage drop between the two leads, respectively.
Here we assume the electron-hole symmetry, which means that the
one-particle energy at the dot site is fixed as $-U/2$, and the inversion
symmetry at the dot site except for the voltage term.
$N_L$ is the sum of the number operators in the left lead
and $N_R$ in the right lead.
In the following we focus on the half-filled case.

In the Keldysh formalism, one starts with an equilibrium state of an
unperturbed Hamiltonian.
One then turns on the perturbation term adiabatically, and gets a
nonequilibrium steady state after a relaxation time.
Although there are several choices for dividing the Hamiltonian into
unperturbed and perturbed terms, the nonequilibrium steady states for
the full Hamiltonian are expected to be independent of the choice.
For the present problem we take a time-dependent voltage term.
$\theta(\tau)$ is a smoothed step function which is introduced to
simulate adiabatic switching-on of the voltage.
Throughout this paper we set $\theta(\tau)$ as
\begin{align}
 \label{qdot_smooth}
 \theta(\tau) \equiv \frac{1}{\exp\left(\frac{\tau_0 - \tau}{\tau_1}\right) + 1}.
\end{align}
The reason for our choice of the time dependence of the Hamiltonian is
that, if $t'$ is chosen as the perturbation, there is only one nonzero
eigenvalue of the reduced density matrix of the lead parts, i.e. the
entanglement entropy between dot and leads are zero for the initial states.
Then the usual DMRG scheme to obtain the optimal basis set, which will
be necessary for the subsequent TdDMRG calculation, cannot be used directly.

By the TdDMRG method we can compute the time evolution of the initial
wave function.
In order to obtain the current at each time we can simply take the
expectation values as follows:
\begin{align}
\label{qdot_JL_JR}
 J(\tau)
 =-\langle \psi(\tau) | e \dot N_L | \psi(\tau) \rangle
 = \frac{iet'}{\hbar} \sum_{\sigma} \langle \psi(\tau) | (c_{0
 \sigma}^{\dagger} c_{-1 \sigma} - h.c.) | \psi(\tau) \rangle.
\end{align}
Note that one can define the current operators for every bond in the system.
When the system reaches a steady state, the currents stay constant at
a certain value because of the definition of the steady states and the charge
conservation.

The steady current in the thermodynamic limit can be
calculated based on the Keldysh formalism as \cite{2nd_order}
\begin{align}
 J(V) = \frac{e}{\hbar} \sum_{\sigma}
 \int_{-\frac{eV}{2}}^{\frac{eV}{2}} d \omega
 \frac{\Gamma_L \Gamma_R}{\Gamma_L + \Gamma_R}
 \left(-\frac{1}{\pi}\right)
 \mathrm{Im} \left( g_{\mathrm{dot}}^{r}(\omega) \right),
\label{qdot_J_Keldysh}
\end{align}
where $\Gamma_{L, R}(\omega)$ is the resonance width due to the
hybridization to the left and right leads which will be assumed to be
the same in this paper, and $g_{\mathrm{dot}}^r(\omega)$ the retarded
Green's function at the dot site.
When $U=0$, the problem is simplified to the one-particle level.
In this case Eq.\eqref{qdot_J_Keldysh} can be exactly evaluated.
Alternatively, one can numerically diagonalize the one-particle
Hamiltonian and explicitly calculate the time-evolution operator for a
finite size system.
Errors included in TdDMRG results can be estimated by comparing with
results obtained from these methods.

Since our system is finite we cannot reach to real steady states.
However, it is possible to evaluate steady currents from the
quasi-steady states which are defined in the time domain after the
initial relaxation time and before the arrival of the wave front
reflected at the boundary of the system.

\begin{figure}[t]
 \begin{center}
  \includegraphics[width=6.8cm]{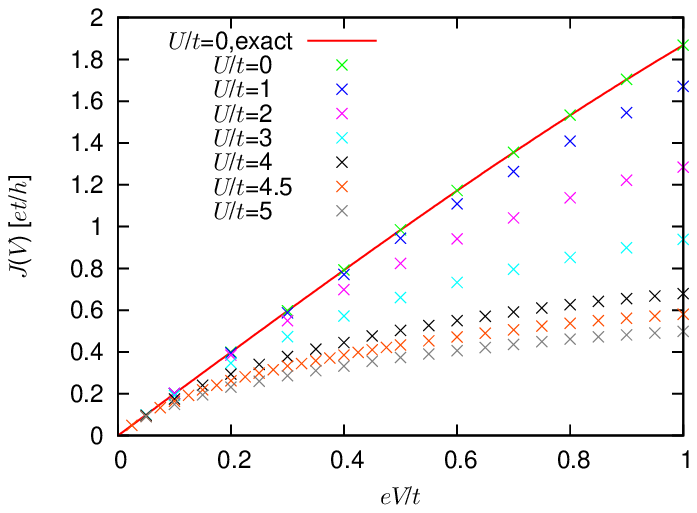}
  \includegraphics[width=6.8cm]{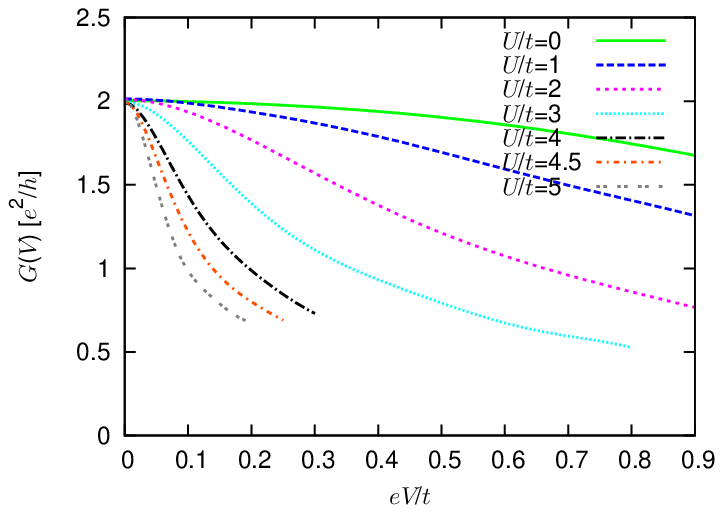}
  \caption{\label{qdot_J-V}
  (Left) Current-voltage characteristics of the Anderson model
  Eq.\eqref{qdot_Hamiltonian}.
  (Right) Differential conductance obtained by interpolating $J(V)$.
  We omit the data where numerical differentiation is so delicate
  that it can produce an unphysical peak structure in $G(V)$.
  Adopted from Ref.\cite{KirinoQD2}.
  }
 \end{center}
\end{figure}

Quasi-steady states can be realized
by the TdDMRG calculation for $U/t \leq 5$.
The fixed value $t'/t=0.6$ and the range for $U$ correspond to the
Wilson ratio $1 \leq R_W \leq 1.97$.
Then we can extract the nonlinear current-voltage characteristics from
the quasi-steady regions, which are summarized in Figs.\ref{qdot_J-V} \cite{KirinoQD2}.
In the noninteracting case the TdDMRG results perfectly agree with the
exact analytic results for $L\rightarrow \infty$.
Moreover, the linear conductance for every $U$ corresponds to the
prediction of the Fermi liquid theory \cite{linear_1},
\begin{align}
 \left. \frac{\partial J}{\partial V} \right|_{V=0} = \frac{2e^2}{h}.
\end{align}

In the differential conductance $G(V)$ we can see a substantial
difference in the width of the zero bias peak depending on $U$ which is
a reflection of the sharp Kondo peak in the local density of states at
the dot site.
Therefore we can say that the TdDMRG calculation catches the essence of
the Kondo effect of the quantum dot system out of equilibrium in
the parameter space investigated.

\section{Serial Double-Quantum-Dot System}
Now we proceed to the serially coupled double-quantum-dot system
\cite{Oosterkamp_dqd,Fujisawa_dqd,Jeong_Science,vandelWiel_review},
each connected with a lead, and discuss transport properties in
the system driven by a DC voltage.
From a theoretical point of view the system can be modeled by the two
impurity Anderson model.

\begin{figure}[b]
 \begin{center}
  \includegraphics[width=6.4cm]{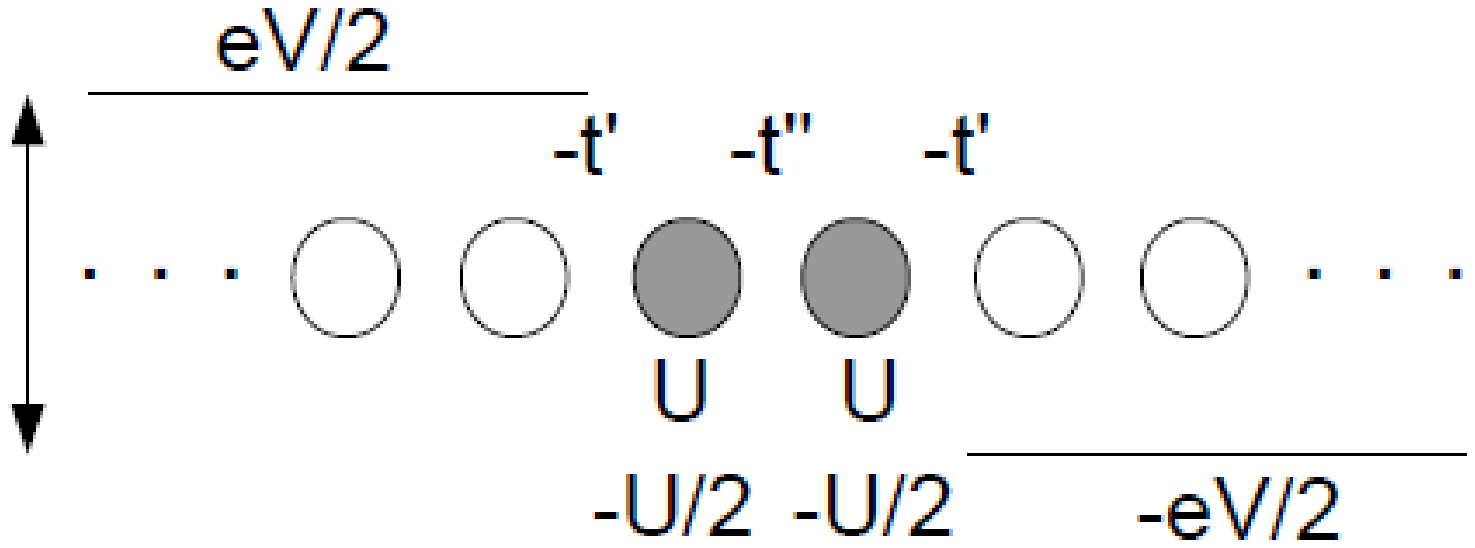}
  \includegraphics[width=4.2cm]{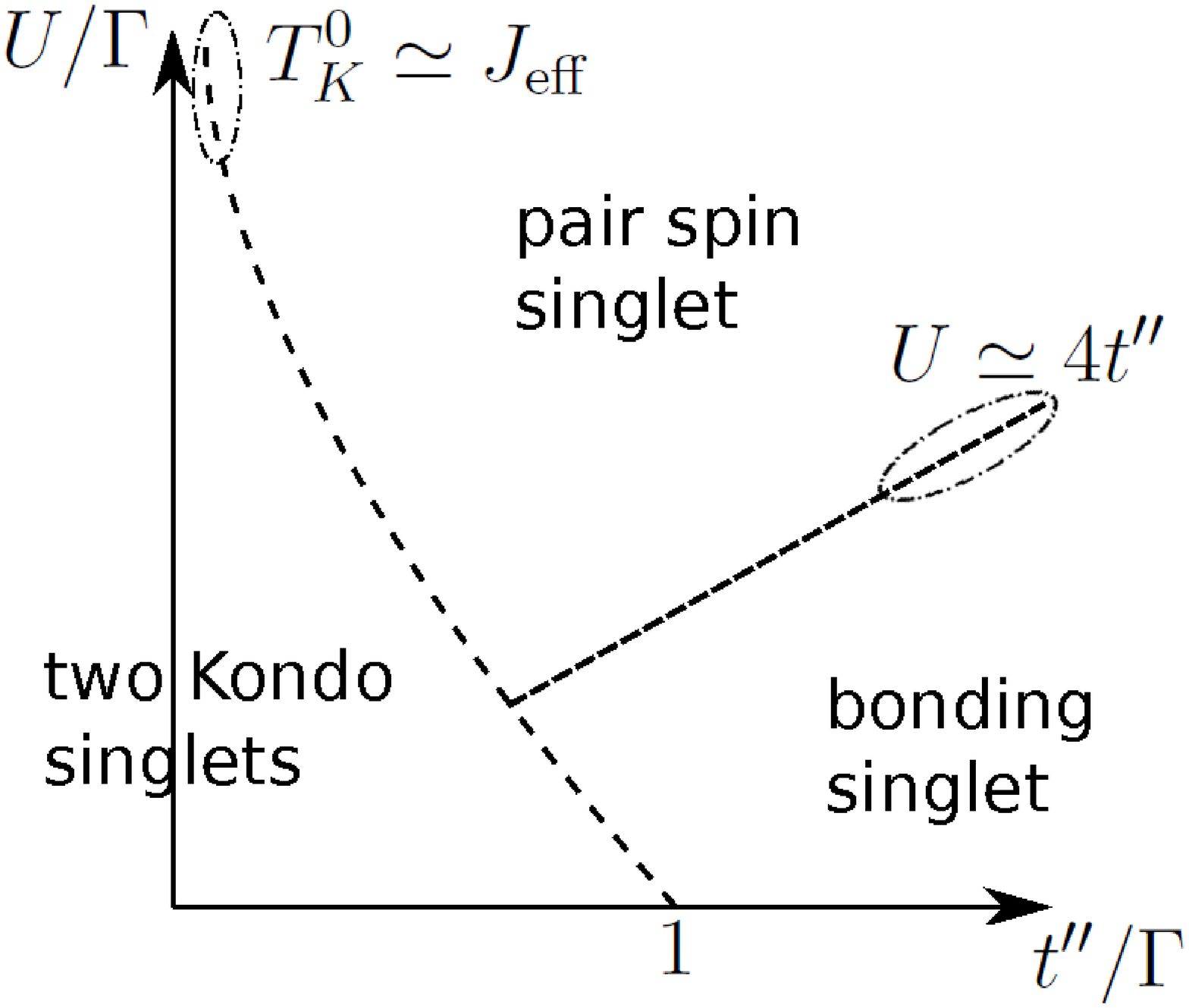}
 \end{center}
 \caption{\label{dqd_fig_Ham}
 (Left) Model of the serial double-quantum-dot system.
 (Right) Schematic diagram for the characteristic ground states (i), (ii) and
 (iii) on the $t''-U$ plane.
 The curves represent crossover between the states.
 }
\end{figure}

The model is depicted in Fig.\ref{dqd_fig_Ham} and it is straightforward
to write the Hamiltonian explicitly in the similar form as
Eq.\eqref{qdot_Hamiltonian}.
The only new additional parameter is the interdot hopping, $-t''$.
In what follows we discuss the current between the two dots,
\begin{align}
 J(\tau) \equiv
 \frac{iet''}{\hbar} \sum_{\sigma} \langle \psi(\tau) |
 (c_{r \sigma}^{\dagger} c_{l \sigma} - h.c.)
 | \psi(\tau) \rangle,
\end{align}
where $c_{l\sigma}^{\dagger}$ ($c_{r\sigma}^{\dagger}$) creates an
electron at the left (right) dot.

There are several theoretical studies on the double-quantum-dot system.
For the linear transport, the slave boson mean field theory
(SBMFT) \cite{GeorgesMeir,AonoEto} and the numerical renormalization group (NRG)
method \cite{IzumidaSakai} have been applied to the problem and predicted
that the linear conductance has a sharp peak
which reaches the perfect conductance $2e^2/h$ as a function of the
ratio of the interdot hopping amplitude $t''$ to the resonance width at
the Fermi energy $\Gamma$, which is given by $\Gamma=t'^2/t$ for the
present model.
The peak is located at $J_{\mathrm{eff}} \sim T_K^0$, where
$J_{\mathrm{eff}} = \frac{4t''^2}{U}$ denotes the antiferromagnetic
exchange coupling between the two dots,
and $T_K^0$ the Kondo temperature at $t''=0$.

\begin{figure}[b]
 \begin{center}
  \includegraphics[width=4.0cm,clip]{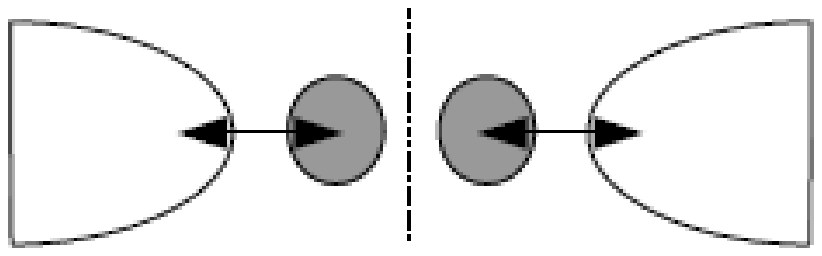}
  \hspace{0.5cm}
  \includegraphics[width=4.0cm,clip]{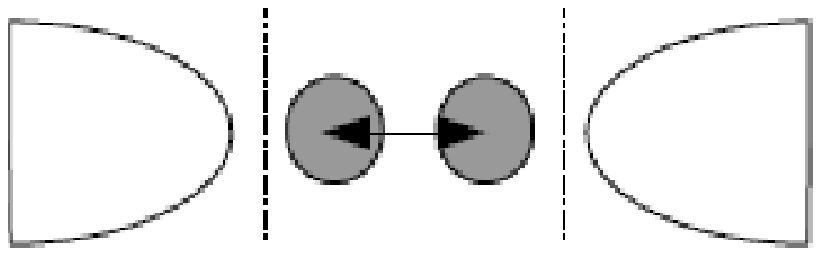}
  \hspace{0.5cm}
  \includegraphics[width=4.0cm,clip]{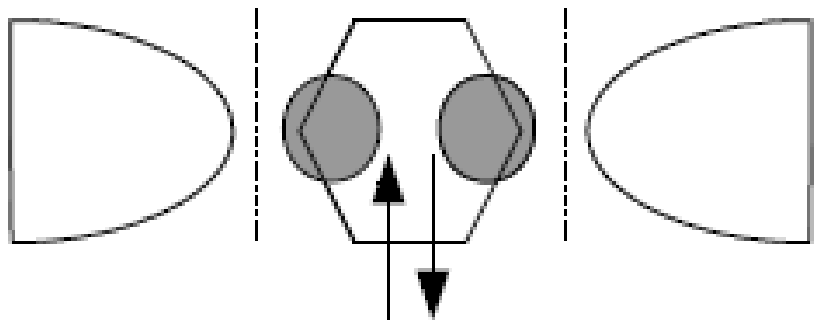}
  \vspace{-0.5cm}
 \end{center}
 \caption{\label{schematic_ground_states}
 (Left)  Schematic figure of the two-Kondo-singlets state.
 (Middle) The pair spin singlet state.
 (Right) The bonding singlet state.
 }
\end{figure}
This behavior can be understood in connection with formation and
competition of the various types of singlets in the system as explained
in Ref.\cite{IzumidaSakai}.
The conductance formula is expressed by \cite{MeirWingreen}
\begin{align}
 G = \frac{2e^2}{h} \sin^2\left(\delta_e - \delta_o\right),
 \label{dqd_conductance}
\end{align}
where $\delta_e$ and $\delta_o$ are the scattering phase shifts for even
and odd channels, respectively.
The even and odd orbitals of the two dots are defined as
$
 c_{e \sigma}^{\dagger} \equiv \frac{1}{\sqrt{2}} \left(c_{l \sigma}^{\dagger} + c_{r
 \sigma}^{\dagger}\right) $ and $
 c_{o \sigma} \equiv \frac{1}{\sqrt{2}} \left(c_{l \sigma}^{\dagger} - c_{r
 \sigma}^{\dagger}\right).
$
When each impurity site contains one electron, $\delta_e+\delta_o=\pi$
follows from the Friedel sum rule.
Then, the following three characteristic states of the two impurity Anderson
model appear depending on $t'', \Gamma$ and $U$:
\begin{itemize}
 \item[(i)] For small $t''/\Gamma$ such that $J_{\mathrm{eff}} \ll T_K^0$, the
	    Kondo singlet state is formed on each dot with its adjacent
	    lead.
	    The phase shifts for this state are
	    $\delta_e \sim \delta_o \sim \pi/2$.
 \item[(ii)] In the intermediate $t''/\Gamma$ region with $J_{\mathrm{eff}} \gg T_K^0$,
	     the local spins on both dots are coupled as a spin singlet
	     state,
	     $\frac{1}{\sqrt{2}} (
	     c_{l \uparrow}^{\dagger} c_{r \downarrow}^{\dagger} -
	     c_{l \downarrow}^{\dagger} c_{r \uparrow}^{\dagger}
	     ) |0\rangle$:
	     a pair spin singlet.
	     $\delta_e \sim \pi, \delta_o \sim 0$.
 \item[(iii)] When $t'' \gtrsim U/4$ the doubly occupied state of bonding
	      orbital of the two dots,
	      $c_{e \uparrow}^{\dagger} c_{e \downarrow}^{\dagger} |0\rangle$,
	      becomes the most stable: a bonding singlet.
	      $\delta_e \sim \pi, \delta_o \sim 0$.
\end{itemize}
The above three states are schematically shown in
Fig.\ref{schematic_ground_states} and their relationship is displayed in
Fig.\ref{dqd_fig_Ham} (b).
For the ground state (i) the two dots are effectively decoupled and
similarly for (ii) and (iii) the two dots can be viewed as detached from the
two leads, resulting in small conductance for all of the above states.
A sharp peak with height $2e^2/h$ is found in the crossover region
between (i) and (ii), namely $J_{\mathrm{eff}} \approx T_K^0$.
The continuous changes of $\delta_e$ and $\delta_o$ from (i) to (ii)
(or directly to (iii) for relatively small $U$) with keeping
the Freedel sum rule ensure that Eq.\eqref{dqd_conductance} has the
maximum which reaches the perfect conductance.
Just at the peak the system forms a coherent superposition of
the two-Kondo singlets on the dots.
As $U$ increases or $\Gamma$ decreases, in other words the electron
correlation becomes strong, $T_K^0$ rapidly decreases and the peak
position moves to smaller values of $t''$.
In addition the width of the peak becomes narrower, representing the
decrease of the Kondo temperature.
These intriguing behaviors can be attributed to the quasi-quantum
criticality of the two impurity Kondo problem.

\begin{figure}[t]
 \begin{center}
  \includegraphics[width=7.2cm]{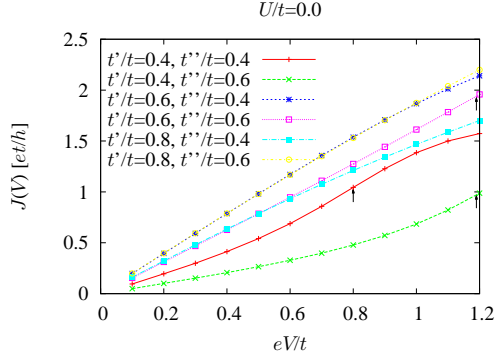}
  \caption{\label{dqd_J-V_0}
  Current-voltage characteristics of the serial double-quantum-dot
  system for $U=0$ obtained by using exact diagonalization.
  Results for $t'/t=0.6$ and $t''/t=0.6$ are obtained for system of $L=81$
  sites and the others are for $L=121$.
  Arrows indicate the positions of the inflection points, that is, the
  peak positions in the differential conductance $dJ(V)/dV$.
  Curves without associated arrows have inflection points at $V=0$.
  }
 \end{center}
\end{figure}

Let us first discuss the current-voltage characteristics in the
noninteracting case.
Since the parameters for the two dots are identical and the energy
levels are set to be particle-hole symmetric, each dot contains
exactly one electron.
Then for $U=0$ the linear conductance is given by
\begin{align}
 G = \frac{2e^2 \Gamma^2}{h} \left| g_e(0+) - g_o(0+) \right|^2
 = \frac{2e^2}{h} \frac{4 \left( t''/\Gamma \right)^2}
 {[1+\left( t''/\Gamma \right)^2]^2}.
\end{align}
$G$ becomes the perfect conductance $2e^2/h$ at the point
$t''/\Gamma=1$.
From Fig.\ref{dqd_J-V_0}, the linear conductance
for $(t'/t, t''/t)=(0.6, 0.4)$ and $(0.8, 0.6)$ is almost the perfect
one ($1.98e^2/h$ and $1.99e^2/h$), since $t''/\Gamma=1.11, 0.94$ is close to unity.

The LDOS for the even and odd states have peaks at
$\pm t''$ with widths $\sim \Gamma$.
The electrons moving through the two quantum dots, which are inside the
bias window $-eV/2 \leq \omega \leq eV/2$, use these peaks in
the resonant tunneling.
When $\Gamma$ is large, the two peaks merge into a single peak located
at $\omega=0$.
On the other hand for $t'' > \Gamma$, the differential conductance
$G(V) = \frac{\partial J(V)}{\partial V}$
has its maximum when the peaks in the LDOS come into the bias window.
Concerning the results shown in Fig.\ref{dqd_J-V_0}, for $t'/t=0.6$ and $0.8$
the maximum of the slope of $J(V)$ is located at $V=0$, reflecting the
single peak structure of the LDOS.
On the other hand for $t'/t=0.4$ we observe the split peaks of $G(V)$ at
$eV=\pm 2t''$.

\begin{figure}[t]
 \begin{center}
  \includegraphics[width=7.2cm]{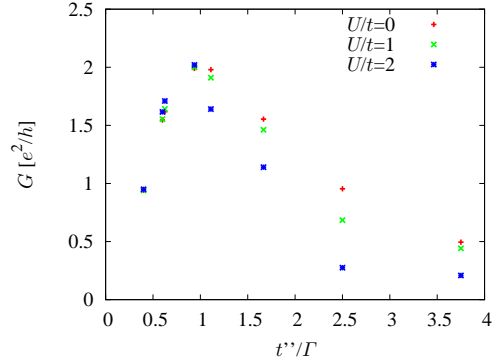}
  \caption{\label{dqd_linear_from_TdDMRG}
  Linear conductance of the serially coupled double-quantum-dot system
  obtained from $J(V)/V|_{eV/t=0.1}$.
  The sets of data are for $(t'/t, t''/t) = (1, 0.4), (1, 0.6), (0.8,
  0.4), (0.8, 0.6), (0.6, 0.4), (0.6, 0.6), (0.4, 0.4)$ and $(0.4, 0.6)$
  from left to right.
  }
 \end{center}
\end{figure}
Now we consider the effects of the Coulomb interaction on the currents.
First we study the linear conductance and it is summarized in Fig.\ref{dqd_linear_from_TdDMRG}.
For each $U$ the conductance as a function of $t''/\Gamma$ forms a peak
near $t''/\Gamma=1$ and their heights reach almost $2e^2/h$
\cite{Al-Hassanieh,GeorgesMeir,AonoEto,IzumidaSakai}.
While on the right side of the peak the conductance decreases as $U$
increases, it increases on the left side, reflecting the shift of the
peak toward smaller $t''/\Gamma$, in accordance with the NRG
results \cite{IzumidaSakai}.

\begin{figure}
 \begin{center}
  \includegraphics[width=7.0cm]{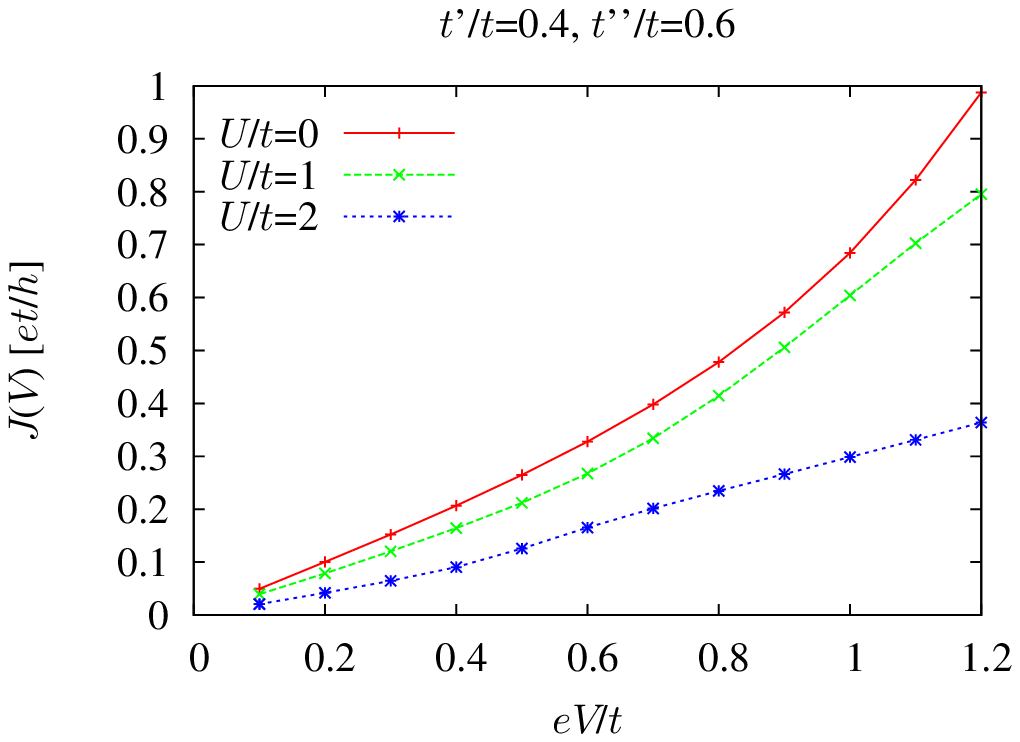}
  \includegraphics[width=7.0cm]{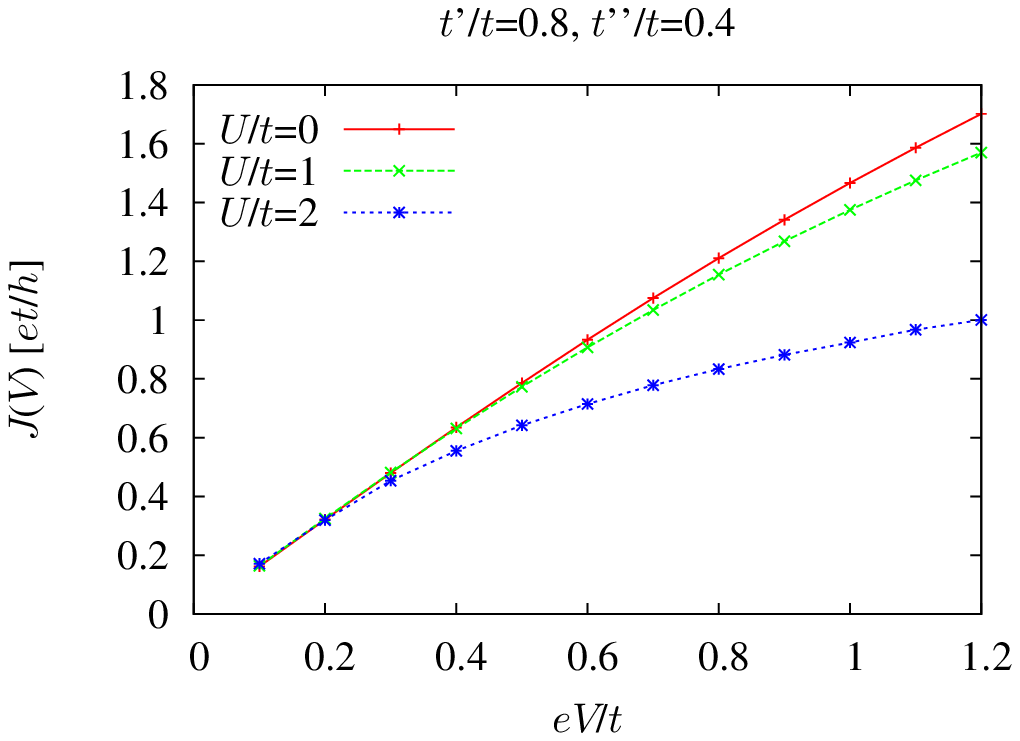}
  \includegraphics[width=7.0cm]{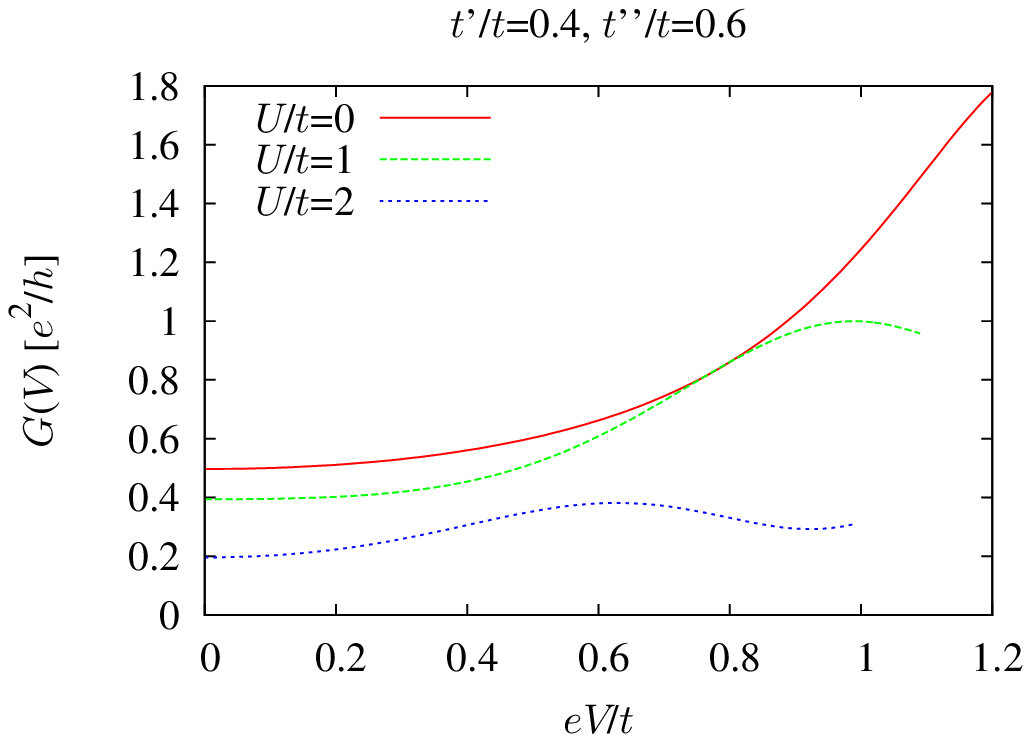}
  \includegraphics[width=7.0cm]{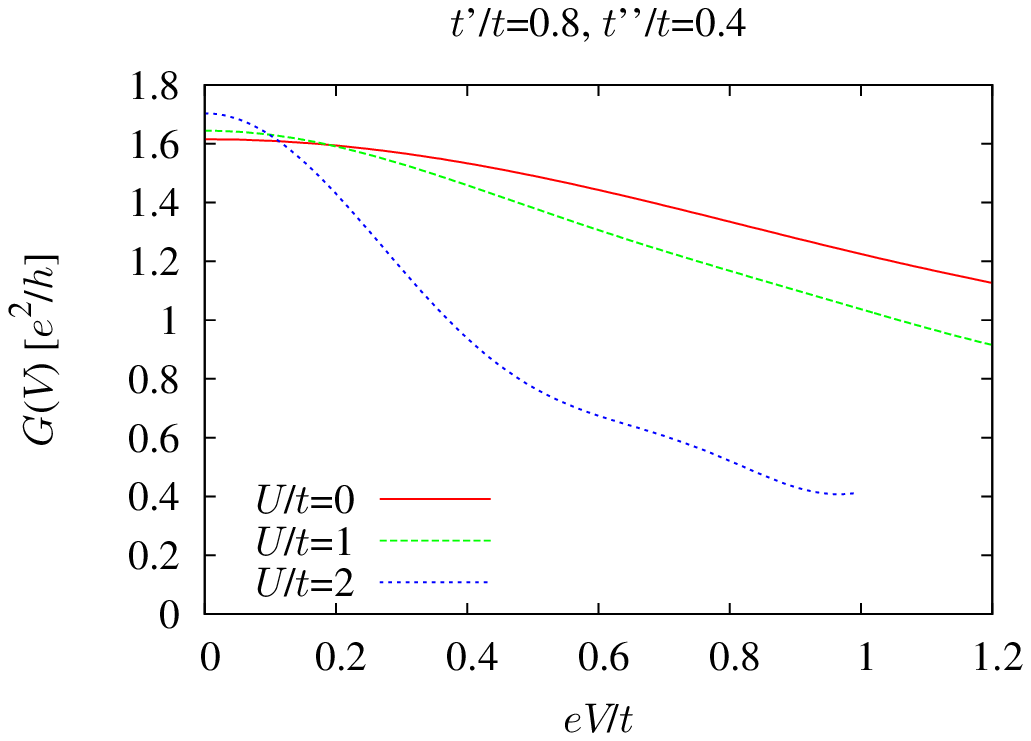}
  \includegraphics[width=7.0cm]{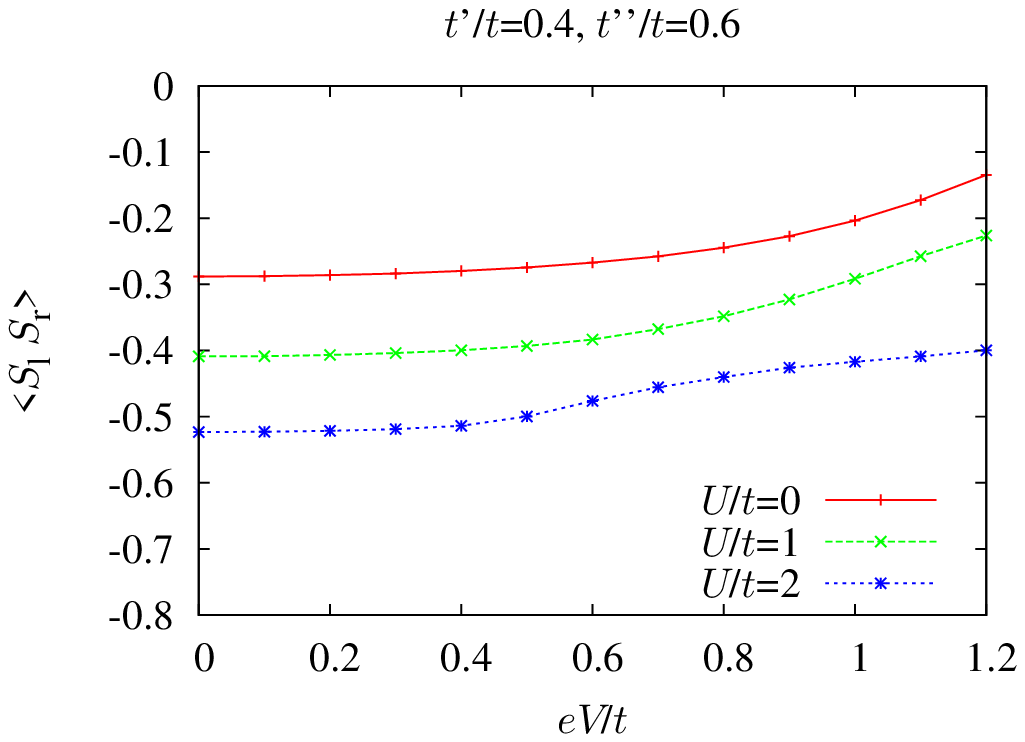}
  \includegraphics[width=7.0cm]{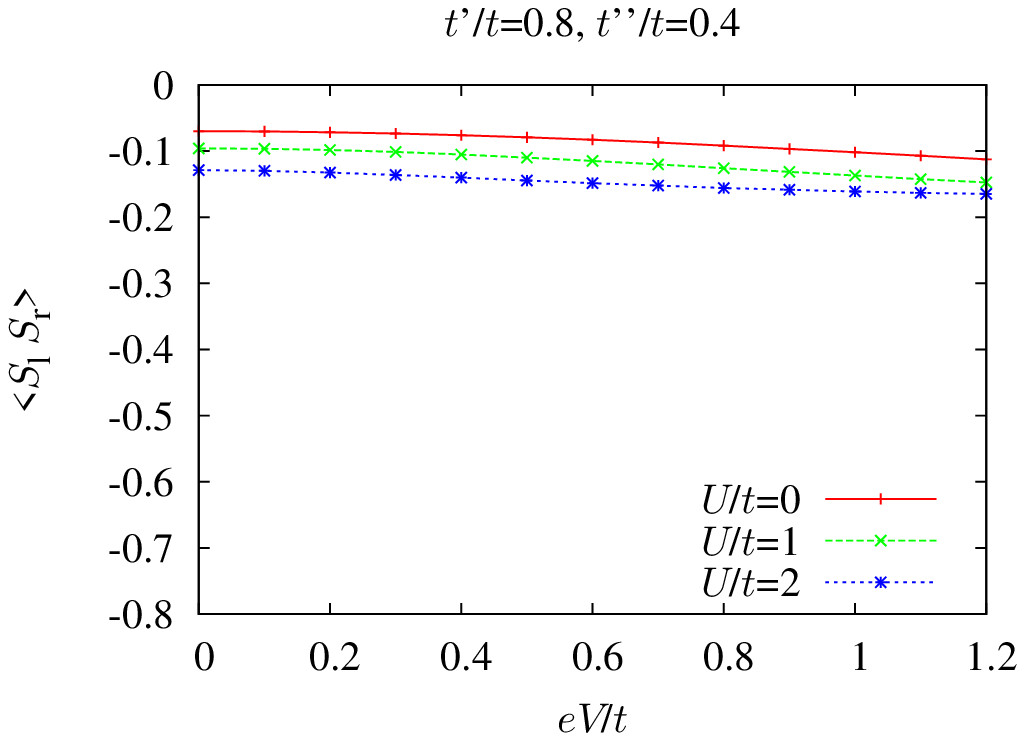}
 \end{center}
 \caption{\label{dqd_fig_J-V}
 Current-voltage characteristics $J(V)$, differential conductance
 $G(V)$ and the expectation values of the spin correlation function
 $\langle \vec S_l \cdot \vec S_r \rangle$ for
 $(t'/t, t''/t)=(0.4, 0.6)$ and $(0.8, 0.4)$.
 The TdDMRG calculations have been performed for the systems with $L=121$.
 }
\end{figure}
Several typical examples of the nonlinear currents of the
double-quantum-dot system are shown in Fig.\ref{dqd_fig_J-V} \cite{PhDThesis},
together with the differential conductance
$G(V)\equiv \frac{\partial J(V)}{\partial V}$ and the expectation values
of the spin correlation operator $\langle \vec S_l \cdot \vec S_r \rangle$.
Note that the results for $G(V)$ have larger errors than the others
since they are obtained via interpolation and differentiation
of the data points of $J(V)$.

The results for $(t'/t, t''/t) = (0.8, 0.6)$, not shown in this paper
\cite{PhDThesis}, are close
to the linear conductance peak at $t''/\Gamma \simeq 1$.
The linear conductance is thus near $2e^2/h$ and the system as a whole
forms a coherent singlet state.
Out of the linear response regime we find a substantial decrease of the
current due to the Coulomb interaction.
This can be attributed to the coherent singlet state whose weight of the
LDOS builds up in a narrow region around the Fermi energy, similarly to
the case of the single dot.

Next we discuss the results on the right side of the linear conductance
peak at $t''/\Gamma \simeq 1$, that is, the results for
$(t'/t, t''/t) = (0.4, 0.6)$, left panels in Fig.\ref{dqd_fig_J-V}.
In this case the ground state has large amplitudes on the states
(ii) and (iii), which can be confirmed by relatively large values of
$|\langle \vec S_l \cdot \vec S_r \rangle|$.
The splitting of the zero bias peak can be seen in $G(V)$ in this case.
The position of the peak $eV_{\mathrm{peak}}$ moves to smaller $V$ as
increasing $U$ and corresponds to the position where
$|\langle \vec S_l \cdot \vec S_r \rangle|$
varies significantly.
This can be explained as follows: as $V$ becomes comparable to $V_{\mathrm{peak}}$
the local electronic state is modified in the steady state.
Hence it is confirmed that the many-body bonding and antibonding states
are located at $\pm eV_{\mathrm{peak}}/2$ in the LDOS.
This allows us to interpret $eV_{\mathrm{peak}}/2$ as the effective
hopping parameter $\tilde t''(U)$.
Our results indicate that $\tilde t''(U)$ decreases by the renormalization
effect due to the interaction.
This simple picture qualitatively agrees with the SBMFT and the NCA
results where the split peaks are located at $eV \ll t''$.
For $(U/t, t'/t, t''/t) = (1, 0.6, 0.6)$ and $(2, 0.6, 0.6)$, not shown
in this paper, the splitting cannot be seen since the relatively large
resonance width $\Gamma$ merges the two peaks at $\pm \tilde t''(U)$
into a single peak.

The SBMFT and the NCA results are for a fixed energy level
in each dot and in the limit $U\rightarrow \infty$.
If we consider the limit of $U\rightarrow \infty$ with keeping
the particle-hole symmetry, we end up with the state (ii),
since $T_K^0 \ll J_{\mathrm{eff}}$ and $t'' \ll U/4$ always hold.
There the pair spin singlet state is frozen and little current can flow.
Thus, judging from our results we expect that the peak position
moves toward $V=0$ and, at the same time, the current is strongly
suppressed by increasing $U$.

Now we turn to the left side of the linear conductance peak.
$G(V)$ for $(t'/t, t''/t) = (0.8, 0.4)$ shows the zero bias peak,
Fig.\ref{dqd_fig_J-V} right panels.
The width of the zero bias peak becomes narrow as increasing $U$,
reflecting the formation of the sharp Kondo resonant peak.
This behavior is basically the same as in the case of the single quantum
dot system.
In the parameter sets considered here, $\Gamma$ is relatively large and
therefore we see a broad linear response regime.
A notable feature in this case is the enhancement of the spin
correlation function $\langle \vec S_l \cdot \vec S_r \rangle$ by the
finite bias voltage.
The antiferromagnetic spin correlation in the steady states turn out to
be enhanced compared with the ground state expectation values.
Note that the states we are focusing on are near the two-Kondo singlet regime
(the state (i)) and the spin correlation is small since the interdot
correlation is effectively suppressed.
By forcing the current flow through the double dots, the interdot
coherence seems to be partly recovered.
This does not conform to a naive expectation that the finite bias
voltage suppresses the electron correlation effects in a similar way as the
finite temperature does.
This enhancement of the interdot correlation by the voltage has similarity
to the suppression of the double occupancy
$\langle n_{\mathrm{dot} \uparrow} n_{\mathrm{dot} \downarrow} \rangle$
by the voltage in the single quantum dot system \cite{KirinoQD,WernerQMC}.

\section{Conclusions}
In this paper we have reviewed recent developments on the TdDMRG method
applied to the nonequilibrium transport through quantum dots.
The results for a single dot show clearly that one can obtain reliable
results on steady currents in systems with correlation effects up to the
intermediate coupling regime.

In the second part of this paper, we have investigated the nonlinear
transport through serially coupled double-quantum-dot system.
We have accurately simulated the quasi-steady states and discussed how
the steady currents are affected by the competition between the
formation of the two-Kondo singlets (the state (i)) and the local
singlet states (the pair spin singlet (ii) and the doubly occupied
bonding singlet (iii)).

It has been found that the differential conductance shows splitting of the zero
bias peak when the ground state is of the local nature, namely near the
state (ii) or (iii).
The positions of the split peaks in the differential
conductance are renormalized to smaller values towards
$0$ as $U$ increases under the condition of the particle-hole symmetry.
In addition we have shown that in the cases near the state (i) the
interdot spin correlation can be enhanced by applying a finite voltage.

The results presented in this paper indicate that it is now possible to
perform unbiased numerical calculations for nonequilibrium steady states
of strongly correlated electron systems.
The approach based on the TdDMRG is from one-dimension and of course a
complementary approach is from higher dimensions, notably from the
infinite dimension.

\begin{acknowledgement}
 It is our great pleasure to dedicate the present paper to Professor
 Vollhardt for his sixtieth birthday.
 This work was supported by JSPS Grant-in-Aid for JSPS Fellows
 21$\cdot$6752, by Grant-in-Aid on Innovative Areas ``Heavy Electrons''
 (No. 20102008) and also by Scientific Research (C) (No. 20540347).
 S. K. is supported by the Japan Society for the Promotion of Science.
\end{acknowledgement}

\end{document}